\documentclass[preprint,3p,times,authoryear]{elsarticle}

\usepackage{amssymb, amsmath}
\usepackage{multirow}
\usepackage{lineno}
\usepackage{hyperref}

\journal{}

\begin{document}
\begin{frontmatter}



\title{Generalization of bibliographic coupling and co-citation using the node split network}

\author[inst1]{Jinhyuk Yun\corref{cor1}}

\affiliation[inst1]{organization={School of AI Convergence, Soongsil University},
            addressline={369 Sangdo-Ro, Dongjak-Gu}, 
            city={Seoul},
            postcode={06978},
            country={South Korea}}
\cortext[cor1]{E-mail address: jinhyuk.yun@ssu.ac.kr}

\begin{abstract}
Bibliographic coupling (BC) and co-citation (CC) are the two most common citation-based coupling measures of similarity between scientific items. One can interpret these measures as second-neighbor relations distinguished by the direction of the citation: BC is a similarity between two citing items, whereas CC is that between two cited items. A previous study proposed a two-layer node split network that can emulate clusters of coupling measures in a computationally efficient manner; however, the lack of intralayer links makes it impossible to obtain exact similarities. Here, we propose novel methods to estimate intralayer similarity on a node split network using personalized PageRank and neural embedding. We demonstrate that the proposed measures are strongly correlated with the coupling measures. Moreover, our proposed method can yield precise similarities between items even if they are distant from each other. We also show that many links with high similarity are missing in the original BC/CC network, which suggests that it is essential to consider long-range similarities. Comparative experiments on global and local edge sampling suggest that local sampling is stable for both similarities in node split networks. This analysis offers valuable insights into the process of searching for significantly related items regarding each coupling measure.
\end{abstract}



\end{frontmatter}


\section{Introduction}
\label{sec:intro}
With the long history of science studies, the investigation of citation-based similarity has been a major topic of research for information scientists \citep{kessler1963bibliographic, kessler1963an, small1973co, marshakova1973system, price2011networks}. Although direct citation (DC) is the clearest and simplest measure of similarity, the linkage may be incomplete because of its innate limitations. For example, authors occasionally omit citations of pertinent sources because of search failure, information overload, or journals’ non-use policies \citep{wilson1995unused}. Measures of relatedness, for example, bibliographic coupling (BC) and co-citation (CC), are used to determine the missing links between two relevant items using shared third items; however, they represent the similarity of different agents and timelines. BC measures the number of common third items in the reference list between two potentially related items; therefore, the similarity is determined by the accumulated backgrounds of the original authors of the target papers prior to publication. Similarly, CC counts the frequency at which two possibly related papers are cited together in common third items. Thus, CC is determined by the cognitive similarity of the descendants. The degrees of similarity determined by the two relatedness measures are correlated, but not identical. There are two viewpoints on this gap between BC and CC. While some argue that there is an absolute degree of accuracy in the similarities \citep{boyack2010co}, others believe that the gap is inevitable because of the nature of citations \citep{glaser2017same}. In other words, the different similarities result from different agents and viewpoints in science and technology, suggesting that the similarities should be selected for analysis purposes.

Conventionally, calculating BC and CC requires the multiplication of an adjacency matrix. The complexity of matrix multiplication is superlinear. Thus, it is quite an expensive task if the dataset is large. Recently, we proposed the node split algorithm \citep{yun2020return}, which can emulate the clustering of BC and CC networks from the structure of a citation network without computing the exact relatedness. The algorithm modifies citation networks that can be considered as bipartite (or two-layer) networks. Citations can be expressed as a network composed of nodes and directed links representing scientific items (papers, patents, \textit{etc.}) and their citations, respectively. DC is a first-neighbor relation between nodes that have direct links between them. Similarly, BC and CC can be considered as second-neighbor relations in which two nodes share common neighbors. From this viewpoint, BC is based on the shared out-neighbors (nodes that have links from a certain node), whereas the CC counts the shared in-neighbors (nodes that have links to a certain node). The node split algorithm eliminates the influence of the first-neighbor relations by splitting a node into two nodes based on the original node’s roles: citing and cited layers. Therefore, building the relatedness network is equivalent to the projection of the node split network; for example, CC networks are equivalent to projecting a citing layer (as virtual links) onto a cited layer. This approach allows us to obtain similar clustering results without incurring a heavy computational load; furthermore, this approach even showed higher consistency with natural language processing-based field-of-study tags than the clusters from the BC and CC networks did \citep{yun2020return}. In short, the node split algorithm with an in-depth understanding of the citation structure can simultaneously enhance accuracy and efficiency.

Although the node split method has shown good performance, the algorithm focuses solely on yielding precise clusters of scientific items. The node split algorithm does not compute the similarity between the items in the same layer, thus reducing the computational cost. However, the usage of CC and BC is not limited to clustering. For instance, relatedness can be used as a proximity measure for paper recommendation \citep{habib2017paper, habib2019sections, heck2011testing},  information searching \citep{tuarob2012improving, eto2013evaluations}, identifying core documents \citep{jarneving2007bibliographic}, discovering missing links \citep{adafre2005discovering}, \textit{etc}. Each of these tasks should be considered as an important subject for clustering; however, the original node split algorithm cannot be utilized for those purposes \citep{yun2020return}. On the other hand, we observed that the cluster node split algorithm conserved information beyond the target relatedness \citep{yun2020return}. In addition, although each relatedness measure is precise from its perspective, these measures still have a limitation of citation-based analysis: the similarity can be calculated if and only if common neighbors exist. However, these can be missing for various reasons, as discussed above \citep{wilson1995unused}. Assume that two non-neighbor papers $A$ and $B$ have never been cited together, but both are frequently cited together with a third paper $C$. The CC similarity between $A$ and $B$ should be $0$; however, there might be an unseen similarity reflected in their shared citations. A similar issue applies to BC. For example, there might be an unmeasured reference similarity between papers $A$ and $B$ if paper $A$ cites paper $C$ and paper $B$ cites paper $D$, where $C$ and $D$ have shared reference $E$. The above relations can be characterized as the fourth-neighbor relations in the citation graph, and there also exist higher-order even-order interactions from shared citations and references for CC and BC, respectively. 

To investigate the above perspectives, we extend our previous study on the node split algorithm \citep{yun2020return} as a backbone structure of similarity; then, we attempt to generalize BC and CC using random walkers on the node split network. In this study, we propose two different methods: i) personalized PageRank (PPR) and ii) neural embedding (EMB). For both methods, the paths between nodes in the same layer (citing or cited layer) must pass the nodes in the other layer. Thus, the similarity network using paths with a length of two should be equivalent to BC and CC networks. PPR directly uses the visiting sequences of random walkers. It measures the probability of visiting a certain node with random walkers starting from a specific node \citep{page1999pagerank}. However, PPR is computationally expensive because it requires repetitive multiplications of an adjacency matrix. Therefore, a low-cost approximation is also required. EMB also uses visiting sequences, but considers the sequences as sentences in language embedding models \citep{peng2021neural}. Using this alternative, we can reduce the computational cost. To understand the characteristics of the proposed measures, we initially focused on the similarity between the target relatedness and those of our proposed methods. We observed a high correlation between them. We also tested the importance of higher-order similarity obtained between distant nodes, which is demonstrated by the distributions of the PPR and EMB similarities by the path length in the node split network. 

Our approach can also be comprehended as a generalized BC and CC that refines the longitudinal coupling \citep{small1997update} and the relative BC/CC \citep{egghe2002co}, which enhances science mapping with indirect connections using temporal groups and lattice properties, respectively. Although non-citation-based similarities have also been used to measure indirect similarities, such as textual similarity \citep{boyack2013improving} and second-order similarity \citep{colliander2012experimental, thijs2013second}, it is difficult to consider them as the fundamental solutions to the problem of missing citation linkages. They are meta-similarities of first-order similarities, essentially subordinate to first-order similarities. Instead, our experiment shows that the proposed measure can capture significant higher-order relations neglected in relatedness networks, based solely on the citation structure itself.

\section{Methods}
\label{sec:met}

\subsection{Node split method for citation network}
\label{subsec:nodesplit}
Citations can be characterized as a network in which nodes and directed links stand for scientific items and citations, respectively. In this network, DC can be interpreted as denoting first-neighbor relations, whereas BC and CC can be understood as denoting second-neighbor relations between items \citep{yun2020return}. Thus, building a CC network is equivalent to a projected citation network that uses cited items as nodes and citing items as virtual links between them. We can also calculate the BC in a reverse manner. Unfortunately, this is practically impossible because a paper can be both a cited and citing node simultaneously. In a previous study, we proposed the node split method for clustering; this method modifies citation networks that can be considered as bipartite networks \citep{yun2020return}. Although the node split method has good accuracy and performance, it is not applicable to other tasks because it does not calculate similarity values. However, the uses of BC and CC are not limited to clustering. Relatedness can be used to search for items \citep{tuarob2012improving} and recommendations \citep{habib2017paper, habib2019sections}, discover missing links \citep{adafre2005discovering}, \textit{etc.} 

\begin{figure*}[!ht]
\centering
\includegraphics[width=\textwidth]{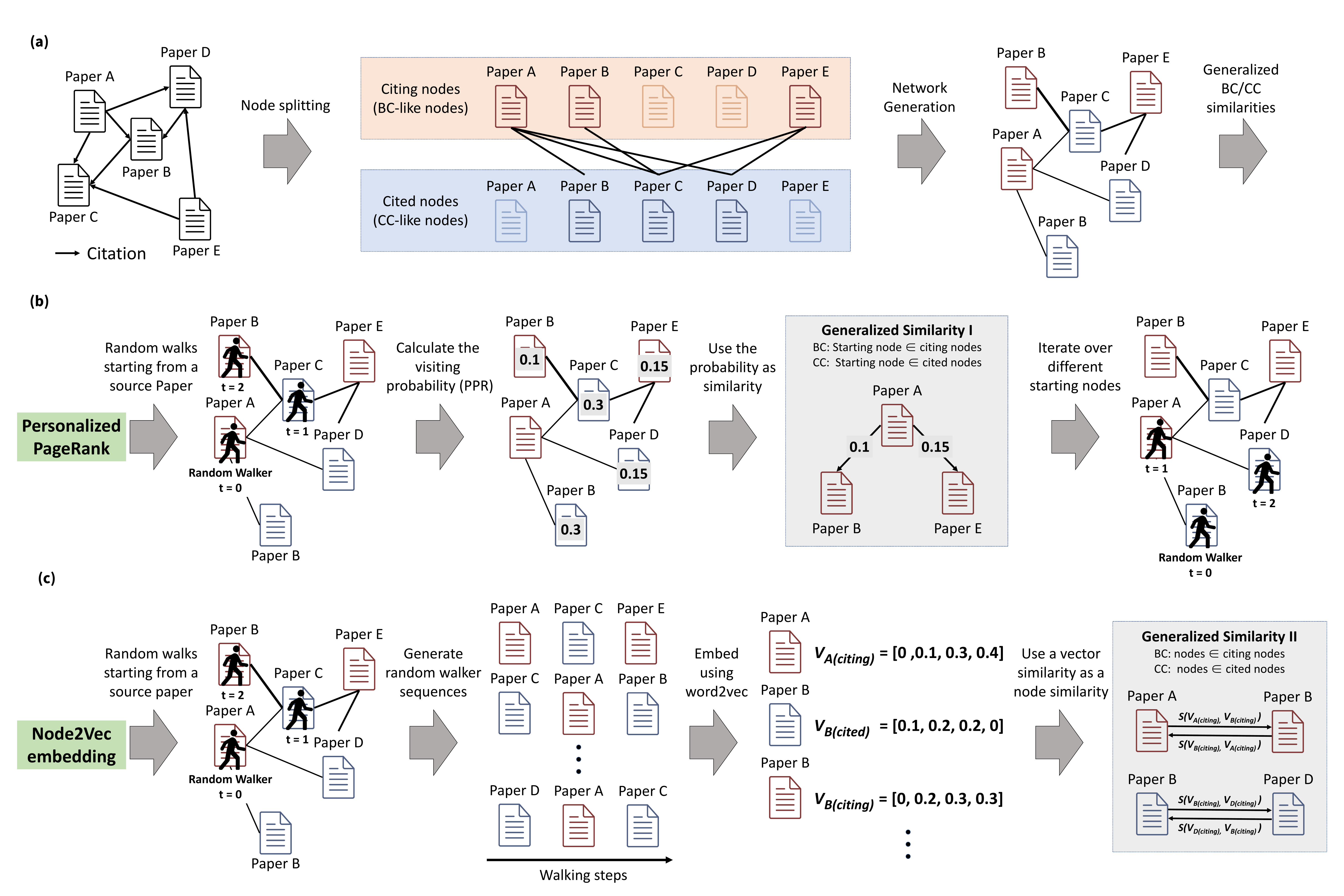}\label{fig:diagram}
\caption{Schematic diagram of generalized CC and BC similarities using a random walker on the node split network. (a) A node split network consists of target articles, and their citations are generated. From the direct citation network, we split a node into two distinct nodes according to the citation directions. We then generate an undirected and unweighted citation network with split nodes and citations. Utilizing the node split network, we obtain two candidates for the generalized BC and CC. (b) We first apply a personalized PageRank algorithm to the node split network \citep{page1999pagerank}. A personalized PageRank between source paper $s$ and target paper $t$ is briefly defined as the probability that a random walker starting from paper $s$ visits paper $t$. Based on the structure of the node split network displayed in (a), the personalized PageRank between citing nodes (PPR) can be used as generalized bibliographic coupling, while those of cited nodes can be used as generalized co-citation. (c) Similarly, the embedding similarity can also be computed using random walkers. From an arbitrary source paper, we generate fixed-length sequences that the walker visits; these sequences are considered as sentences in the word2vec model \citep{peng2021neural}. We then apply neural embedding to the papers to obtain the representation vector for each paper. The vector similarities between citing nodes can also be used as generalized BCs, whereas those of cited nodes can be used as generalized CCs.}
\end{figure*}

Therefore, we propose two possible extensions of the node split method that allow the node split network to compute the similarity between items. We first split each item $P(i)$ into two items based on the original item’s roles in the same way as in the previous study \citep{yun2020return}: a citing node $P_o(i)$ and cited node $P_i(i)$ (Fig.~\ref{fig:diagram}(a)). When a citation exists in the DC network, the two items in different layers are linked while avoiding direct links in the same layer. We then remove isolated nodes with no links and convert all links into undirected ones. The resulting network is the so-called raw node split network. Then, we tested two random-walker-based methods to calculate the similarity between items on the node split network: personalized PageRank (PPR; Section~\ref{subsec:ppr}) and neural embedding (EMB; Section~\ref{subsec:emb}).

\subsection{Personalized PageRank as generalized relatedness}
\label{subsec:ppr}
Personalized PageRank (PPR), also known as random walks with restarts, is a widely used node similarity measure in network analysis \citep{page1999pagerank, haveliwala2003topic}. For a given source $s$, target $t$, and teleport probability $\alpha$, the PPR similarity between two nodes represents the probability that a random walker starting from $s$ will visit $t$. 

We consider the node split network between scientific items, where each node is split into a citing and cited node. There is a link between the citing node of paper $A$ and the cited node of paper $B$ on the node split network if paper $A$ cites paper $B$. On the node split network, a random walker begins to walk from a given source node $s$, moving along the links with equal probabilities for each neighbor (Fig.~\ref{fig:diagram}(b)). During this traversal, the walker occasionally returns to the source node $s$ by the teleport probability $0\leq\alpha\leq1$. The walker also returns to the source node $s$ if the current location of the walker has no links; however, this should not happen for an undirected or bidirectional network because the walker can always return to the preceding node. Practically, PPR can be calculated by recursive multiplication of matrix $M$, which is defined as

\begin{equation}
r = M \cdot r.
\end{equation}

\noindent Here, $M$ is a dense matrix in which an element $M_{ij}$ indicates PPR similarity from node $i$ to node $j$. The elements $M_{ij}$ are defined as follows:

\begin{equation}
M_{ij} = 
\begin{cases}
(1-\alpha)M'_{ij} + \alpha, & \text{if } i = s \\
(1-\alpha)M'_{ij}, & \text{otehrwise}
\end{cases}
\end{equation}

\noindent where $M'_{ij} = A_{ij}$ at the beginning of recursive multiplications. We use the teleport probability $\alpha = 0.15$ following the original study \citep{page1999pagerank}. We then filter the target nodes ${t}$ of the same type (citing or cited) as the source $s$ to yield the similarities. If a source $s$ is a citing node, the similarity should be associated with the BC because a citing node emulates bibliographic coupling \citep{yun2020return}. Similarly, we can yield a proxy for CC when a source node $s$ is a cited node. We then continue iterating over a different source node $s$ until each node has been used as a source node. Note that it is impossible to obtain PPR similarities if there is no path between the two nodes. In this study, we compute PPR using the Python-based \texttt{NetworkX} package \citep{hagberg2008exploring}.

\subsection{Neural embedding as the generalized relatedness}
\label{subsec:emb}
As we already discussed for PPR, we hypothesize that the trail of random walkers on the node split network can capture the semantic similarity between two nodes. Because citing and cited nodes act as nodes in the BC and CC networks, respectively, this method may capture the similarity of the BC and CC networks. Therefore, PPR is expected to emulate BC and CC at least theoretically; however, it requires recursive multiplication of a large and dense matrix $M$; thus, the computation is intensive. To overcome the above difficulties, we also propose an alternative approach for yielding the similarity between two nodes in the node split network, which is called neural embedding similarity (EMB). Graph embedding is a task that finds the vector representation of nodes in which each vector is associated with the nodes and their relations (i.e., links). If the embedding is appropriate, the similarity of the embedded vectors between two nodes should reflect their similarity in the original network; thus, it can be used as an alternative similarity between nodes. We used an embedding model based on DeepWalk and node2vec \citep{perozzi2014deepwalk, grover2016node2vec}, using the trails of random walkers on the network.

We again consider the node split network between papers. First, we generate multiple citation trails $\{T\}$ per source node for all nodes in the node split network. Because there is no intralayer link, the trails are alternately cited nodes and citing nodes, as follows:

\begin{equation}
T_{s;i} = 
\begin{cases}
[P_{cited}(s), P_{citing}(2), P_{cited}(3), P_{citing}(4), ..., P_{cited}(n-1), P_{citing}(n)]  & \text{if } s \text{ is a cited node}, \\
[P_{citing}(s), P_{cited}(2), P_{citing}(3), P_{cited}(4), ..., P_{citing}(n-1), P_{cited}(n)]  & \text{if } s \text{ is a citing node}.
\end{cases}
\end{equation}

Thus, the influence of the DC should be limited for the similarities between nodes in the same layer. We generate $20$ citation trails per source node; these trails stop after $20$ hops. We then train $50$-dimensional vectors using a skip-gram with a negative sampling model. Similar to the conventional word2vec model, similar papers are located near each other in vector-space representations \citep{peng2021neural}; thus, we compute the cosine similarity between all node pairs as a proxy of semantic similarity between papers in the BC and CC networks. We train the neural embedding model using the \texttt{Gensim} package \citep{rehurek2010software}. Unlike the PPR similarity, it is possible to obtain the EMB similarities even if there is no path between two nodes; therefore, this could be a more general approach compared to PPR.

\subsection{Building test networks}
\label{subsec:benchmark}
Although the proposed method can theoretically emulate citation-based relatedness, it should be verified using empirical citation data. Thus, we collected a test network using the February 15, 2021, dump of the Microsoft Academic Graph, which contains 252,109,820 scientific items from the Microsoft Academic Service in \texttt{TSV} format. These data are composed of scientific items such as journal papers, patents, conference proceedings, books, \textit{etc.}, along with citations between them \citep{sinha2015overview, wang2019review}. All items are tagged with six levels based on their fields of study (FOSs, from Level 0 to Level 5) assigned by MAG using natural language processing algorithms. We first gathered papers belonging to the following FOSs for the test: knowledge graphs (Level 3), graph theory (Level 3), and complex systems (Level 2). We also collected articles published in three prominent information science journals: the Journal of Informetrics (JOI), the Journal of the Association for Information Science and Technology (JASIST), and Scientometrics. For all test sets, we used all items regardless of document type (DocType).

\begin{table}[h]
\centering
\caption{Total number of nodes and links for each sampled networks.}\label{tab:stat}
\begin{tabular}{llllll}
\hline
\textbf{Network type} & \textbf{Statistics} & \textbf{Information science} & \textbf{Knowledge graph} & \textbf{Graph theory} & \textbf{Complex system} \\ \hline
\textbf{DC} & Nodes & 11211 & 3375 & 537 & 28529 \\
 & Links & 76102 & 11391 & 922 & 49552 \\
\textbf{DC (GCC)} & Nodes & 11036 & 3160 & 473 & 21106 \\
 & Links & 75996 & 11265 & 881 & 44220 \\
\hline
\textbf{BC} & Nodes & 9371 & 2441 & 385 & 16009 \\
 & Links & 1964258 & 900150 & 47908 & 1120910 \\
\textbf{BC (GCC)} & Nodes & 9334 & 2391 & 377 & 14974 \\
 & Links & 1964206 & 900076 & 47892 & 1118112 \\
\hline
\textbf{CC} & Nodes & 8908 & 1616 & 203 & 10613 \\
 & Links & 888130 & 49714 & 1500 & 136688 \\
\textbf{CC (GCC)} & Nodes & 8866 & 1565 & 193 & 9676 \\
 & Links & 888066 & 49648 & 1490 & 134478\\
\hline
\end{tabular}
\end{table}

We then filtered the citation relations so that both citing and cited articles belonged to the target subset without citation windows. The sets ranged from containing $537$ nodes with $922$ links (graph theory) to $28\,529$ nodes with $49\,952$ links (complex systems; see Table \ref{tab:stat}). We then built networks of the original BC and CC using matrix multiplication of the adjacency matrix ($C = AA^T$ for CC and $B = A^TA$ for BC); thus, the raw coupling strength was defined as the number of overlapped citations (for CC) or overlapped references (for BC). We accounted for only the giant connected component (GCC) of each network, and thus, the effective numbers of nodes were smaller than those of the original sets (detailed statistics are in Table~\ref{tab:stat}). Unlike in our previous study \citep{yun2020return}, the number of nodes in the GCC should be the same for both the node split network and its counterparts because no link filtering was applied. Specifically, the number of nodes in the BC network should be the same as the number of citing nodes in the node split network. This is also true for the nodes in the CC network and the cited nodes of the node split network. An important merit of the proposed method is that the similarity between nodes can be computed even if there is no direct coupling between the nodes. Therefore, it can measure the degree of relatedness of any pair in the subset. Note that EMB can estimate the similarity between two nodes that are not in the same component; however, PPR cannot.

We then applied up-to-date egocentric link normalization for the original BC and CC, which divides the link weights by the total sum of weights for a paper P(i) \citep{waltman2012new}; therefore, the normalized link weight for the BC and CC of nodes $P(i)$ to $P(j)$ is defined as follows:

\begin{equation}
    \widetilde{r}_{ij} = \frac{r_ij}{\sum_{k}{r_ik}}.
\end{equation}

Here, $r_ij$ is the raw strength of the relatedness between nodes $P(i)$ and $P(J)$. Therefore, the link strengths for all items were set to the same magnitude of $1$. Similarly, the sum of all out-link weights for PPR should be $1$ based on its definition as a probability, but not for EMB. We did not apply any normalization to EMB because there might be negative link weights in the EMB networks; however, we did not observe this in our analysis.

\section{Results}
\label{sec:results}
Our primary goal was to emulate the original relationship similarity through a node split network using random walkers. For our analysis, we used four subsets of citation networks from the February 15, 2021 dump of the Microsoft Academic Graph (see Methods). To proceed with the detailed analysis of semantic similarity, we stress the fact that PPR is more computationally expensive than EMB and the original relatedness. PPR requires repeated multiplication of the adjacency matrix, whereas CC and BC only perform a single multiplication. Our motive was to validate a possible generalization of the relatedness, and not to compute the complete similarity between all scientific items; therefore, we chose four scaled-down subsets (see Table~\ref{tab:stat} and Section~\ref{subsec:benchmark}).

\subsection{Similarity between the weights of a relatedness network and generalized networks}
\label{subsec:similarity}
Here, we present evidence that both of the proposed methods can reproduce the original relatedness. We primarily focus on the fact that our method provides similarities closely correlated to the target relatedness. Because the similarities are based on citations, they should reflect the structure of science and technology. In other words, the similarities between different methods should be associated with other similarities to some degree. Thus, we compared the correlation between relatedness and its generalization with the baseline correlation measured between BC and CC. Note that the generalized similarity measures, \textit{i.e.,} PPR and EMB, can estimate the similarity even if the nodes are distant, while BC and CC can only measure similarity when two articles share references or citation sources, respectively. Therefore, we first focus on the similarity between two node pairs in which the original relatedness exists.

\begin{figure*}[!ht]
\centering
\includegraphics[width=\textwidth]{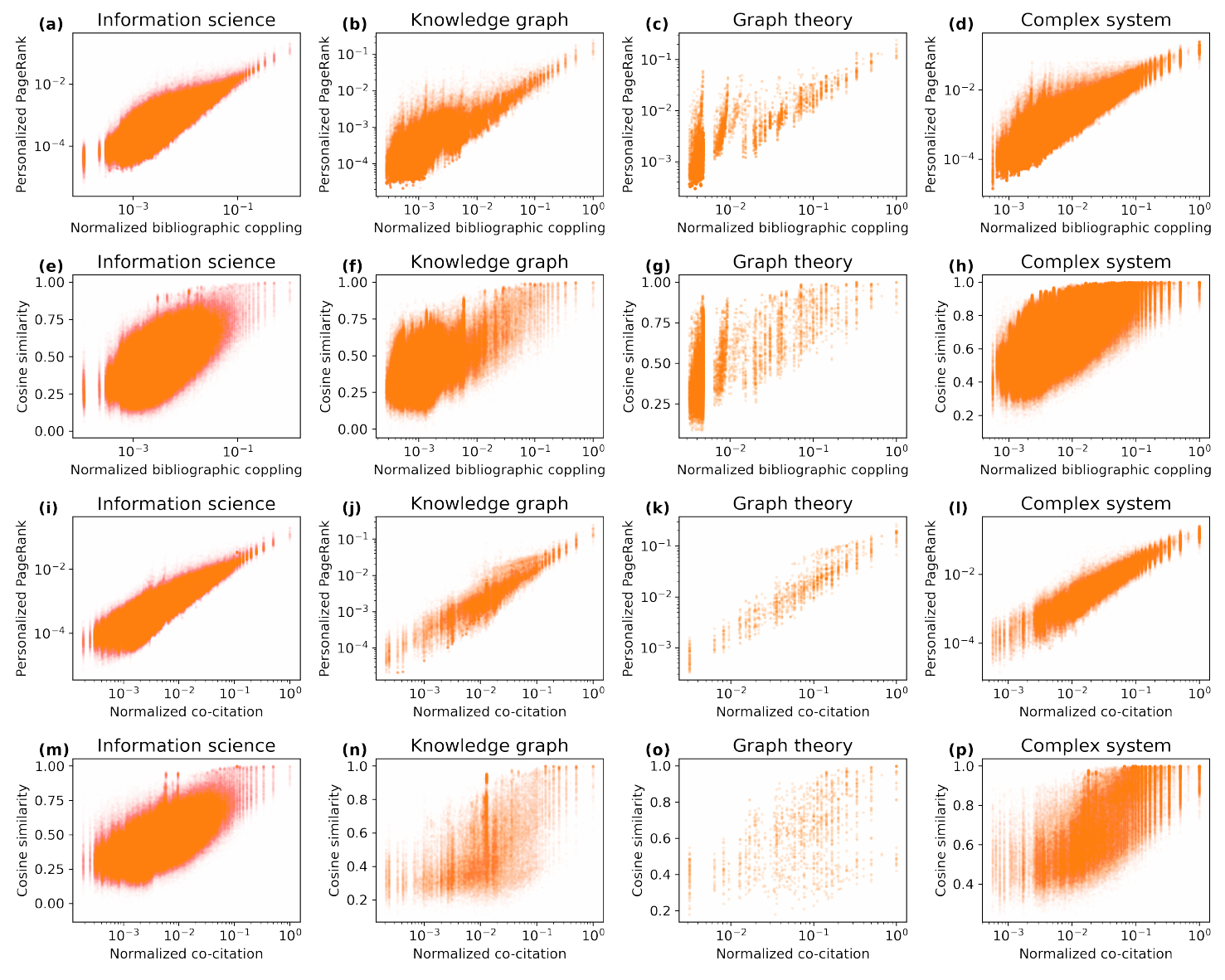}
\caption{Correlation between the coupling measures and their possible generalizations for four different disciplines: (a--d) BC and PPR, (e--h) BC and EMB, (i--j) CC and PPR, and (m--p) CC and EMB.}\label{fig:scatterplot}
\end{figure*}

\begin{table}[]
\centering
\caption{Correlation between generalized relatedness and their corresponding original relatedness.}\label{tab:corr}
\begin{tabular}{lllllll}
\hline
\textbf{Correlation} & \textbf{\begin{tabular}[c]{@{}l@{}}Original\\ similarity\end{tabular}} & \textbf{\begin{tabular}[c]{@{}l@{}}Generalized\\ similarity\end{tabular}} & \textbf{\begin{tabular}[c]{@{}l@{}}Information\\ science\end{tabular}} & \textbf{\begin{tabular}[c]{@{}l@{}}Knowledge\\ graph\end{tabular}} & \textbf{\begin{tabular}[c]{@{}l@{}}Graph\\ theory\end{tabular}} & \textbf{\begin{tabular}[c]{@{}l@{}}Complex\\ system\end{tabular}} \\ 
\hline
\textbf{Pearson} & BC & PPR & 0.8996 & 0.9171 & 0.8939 & 0.9223 \\
 & BC & EMB & 0.3283 & 0.2820 & 0.2466 & 0.3362 \\
\textbf{} & log(BC) & EMB & 0.5319 & 0.3861 & 0.5470 & 0.5996 \\
 & CC & PPR & 0.9219 & 0.9233 & 0.8876 & 0.9314 \\
\textbf{} & CC & EMB & 0.4351 & 0.3946 & 0.3506 & 0.4882 \\
 & log(CC) & EMB & 0.6288 & 0.5126 & 0.4924 & 0.7016 \\
\textbf{} & BC & CC & 0.3106 & 0.2461 & 0.0129 & 0.3478 \\ \hline
\textbf{Spearman} & BC & PPR & 0.7526 & 0.6860 & 0.6568 & 0.8551 \\
\textbf{} & BC & EMB & 0.5942 & 0.5149 & 0.3788 & 0.6282 \\
 & CC & PPR & 0.8667 & 0.8994 & 0.9408 & 0.9430 \\
\textbf{} & CC & EMB & 0.6516 & 0.5342 & 0.5129 & 0.6917 \\
 & BC & CC & 0.3891 & 0.1378 & 0.0223 & 0.5687 \\ 
\hline
\end{tabular}
\end{table}

We found that both PPR and EMB seemed to have a positive correlation for all subsets, despite slight differences in detail (Figure~\ref{fig:scatterplot}). First, PPR showed a strong and positive correlation with BC and CC in Figure~\ref{fig:scatterplot}(a--d) and (i--l). For example, the interrelation between the BC and PPR was nearly linear with respect to the Pearson correlation (see Table~\ref{tab:corr}): $\simeq 0.8996$ for information science (Figure~\ref{fig:scatterplot}(a)), $\simeq 0.9171$ for knowledge graphs (Figure~\ref{fig:scatterplot}(b)), $\simeq 0.8939$ for graph theory (Figure~\ref{fig:scatterplot}(c)), and $\simeq 0.8996$ for complex systems (Figure~\ref{fig:scatterplot}(d)). Similarly, there was also a high Pearson correlation between CC and PPR, as follows: $\simeq 0.9219$ for information science (Figure~\ref{fig:scatterplot}(i)), $\simeq 0.9233$ for knowledge graphs (Figure~\ref{fig:scatterplot}(j)), $\simeq 0.8876$ for graph theory (Figure~\ref{fig:scatterplot}(k)), and $\simeq 0.93 14$ for complex systems (Figure~\ref{fig:scatterplot}(l)). The outliers can have a significant impact on the Pearson correlation \citep{kim2015instability}. One may argue that the Pearson correlation may be inaccurate because citations are known to frequently produce a heavy-tailed distribution with many outliers. However, this trend is still valid for the Spearman rank correlation, which is known to be more stable with regard to outliers. Specifically, the correlations were $>0.65$ and $>0.86$ for citing and cited nodes, respectively, for all subsets. These correlations were much higher than the baseline computed between BC and CC (Table~\ref{tab:corr}). 

\begin{figure*}[!ht]
\centering
\includegraphics[width=\textwidth]{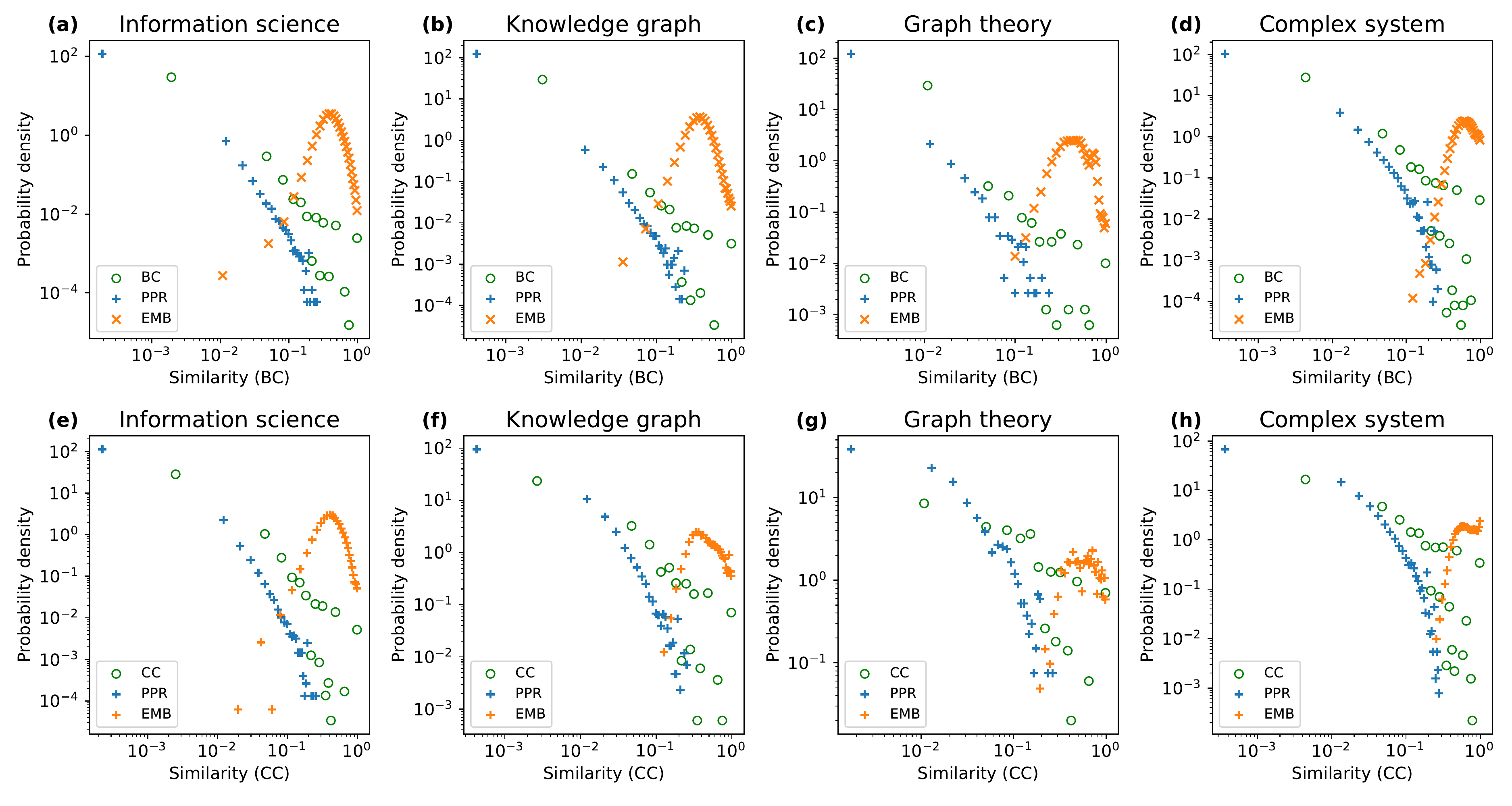}
\caption{The probability density of the original relatedness measures and their generalization: (a--d) BC with its counterparts and (e--h) CC with its counterparts. For comparison, we sampled the node pairs with a direct link in the original relatedness (BC/CC) network; thus, distant node pairs with shortest path lengths of $>2$ in the relatedness network are neglected in the computation of probability density. For the distributions of distant node pairs, see Figure~\ref{fig:dist_by_pathlen}.}\label{fig:distribution}
\end{figure*}

However, the results are unclear when we apply EMB to generalized relatedness. Although a positive correlation can be observed between EMB and the original relatedness, the degree of correlation is insufficient to find the analogous relationship. For instance, the Pearson correlation between BC and EMB for the information science papers is only $\simeq 0.3283$, whereas that between BC and CC is $\simeq 0.3106$ (Table~\ref{tab:corr}). The correlation between BC and EMB ($\simeq 0.3362$) is even smaller than the BC-CC correlation ($\simeq 0.3478$) with respect to complex systems. The interrelations between CC and EMB are superior to those of BC and CC, as reflected by the Pearson correlations, yet they are only $<0.5$, which makes it difficult to conclude that they are connected. We suspect that the lower correlation might be due to the difference in the shape of the distribution; for instance, the outliers may lead to inaccuracy in the correlation analysis, as mentioned above \citep{kim2015instability}. The distribution of PPR and relatedness seems to be skewed and heavy-tailed, whereas EMB follows a centered distribution for all four subsets (see Figure~\ref{fig:distribution}); therefore, the Pearson correlation is not a proper measure of relatedness. We assume that a suitable compensation of outliers may increase the correlation based on their distributions. Indeed, the correlation with EMB always outperformed the raw BC when we took the logarithm of BC. For example, the correlation of EMB with $\log(BC)$ is  $0.5319$ for information science papers, which is much higher than that with raw BC (see Table~\ref{tab:corr}). For the same pairs, the log-relatedness always had a higher correlation with the EMB measure than with the raw relatedness. This is also supported by the strong positive Spearman rank correlation for EMB (Table~\ref{tab:corr}). However, the correlation of EMB was consistently lower than that of \ PPR.

In summary, PPR successfully emulates the target relatedness. Despite its lower correlation in the results, EMB can also be considered as an appropriate alternative for relatedness because of its high correlation with rank. Hence, we use both candidates as possible generalizations of relatedness in the following analysis. 

\subsection{Comparison of computational complexity between the original and generalized relatedness measures}
\label{subsec:complexity}
One reason that we propose EMB for generalization is that it has advantages in terms of computational cost, despite its lower accuracy compared to PPR. EMB is less expensive than PPR in terms of computation. Calculating BC and CC is costly. For instance, the CC matrix can be obtained by multiplying the adjacency matrix by its transpose, which can be written as $C = AA^T$. The BC matrix is also given by $B = A^TA$. Here, $A$ is an $m \times m$ citation adjacency matrix, and $m$ is the number of nodes in the dataset. The complexity of matrix multiplication can be optimized up to $O(nnz(A) \times m)$ using sparse matrix algorithms, where $nnz(A)$ is the number of nonzero elements for an adjacency matrix $A$. For the citation network, $nnz(A)$ is the same as the total number of citations in the dataset. Computing PPR is even more computationally expensive than computing BC and CC. Practically, PPR is estimated by recursive multiplication of matrix $M$, which is a dense matrix in which an element $M_{ij}$ indicates the PPR similarity from node $i$ to node $j$ as follows:

\begin{equation}
r = M \cdot r.
\end{equation}

\noindent Here, $M$ is a dense matrix in which an element $M_{ij}$ indicates the PPR similarity from node $i$ to node $j$. The multiplication is performed recursively until the matrix converges. For a source node $s$, an element $M_{ij}$ is given by $(1-\alpha)M'_{ij} + \alpha$, where $M'_{ij}=A_{ij}$ at the beginning of the overall iteration process and $\alpha$ is the probability that a random walker returns to the source (see \ref{subsec:ppr}); otherwise, an element $M_{ij}$ for all other nodes is $(1-\alpha)M'_{ij}$. Here, $M$ is an $m \times m$ matrix, and $m$ is the number of nodes in the dataset. Thus, the computational complexity is up to $O(k \times nnz(M) \times m)$, where $k$ is the number of iterations. Practically, the number of zero elements of $M$ approaches $0$. Therefore, the computational complexity is approximately $O(k \cdot m^3)$, which is at least k times larger than that of BC and CC.

In contrast, neural embedding can be a lightweight alternative in terms of computation. There are two steps for EMB: i) neural embedding and ii) calculation of the similarity using the embedded vectors. First, the embedding requires $O(N * log(m))$ computations, where $N = l \times n$ is the total size of randomly generated sequences using the skip-gram model with negative sampling. Here, $l$ is the fixed length of the random walker sequence ($l=20$ in this study), and $n$ is the number of random walkers. Because both $l$ and $n$ are tunable parameters, we can compromise between the computational cost and the embedding accuracy. We then compute the cosine similarity for all possible node pairs. This process requires the matrix multiplication of $EE^T$, where $E$ is the $m \times d$ embedded vector; therefore, its time complexity is $O(d \cdot nnz(E))$. Taken together, the total complexity of the neural embedding should be $O(d \cdot m^2)$ because the embedding matrix $E$ is dense.

In summary, the calculation of the PPR of $O(k \cdot m^3)$ is slower than both the EMB of $O(d \cdot m^2)$ and the original relatedness of $O(m \cdot nnz(A))$. A comparison between EMB and relatedness should consider the sparsity of the adjacency matrix. In general, the citation relationships are sparse; thus, $nnz(A) \ll m^2$. Therefore, the computation of EMB is greater than that of the original relatedness. Despite the high computational cost, it should be noted that PPR and EMB can yield similarities between distant nodes in the original relatedness network, which cannot be calculated using the original relatedness.

\subsection{Missing significant links in the network of the original relatedness}\label{subsec:missing_links}

\begin{figure*}[!ht]
\centering
\includegraphics[width=\textwidth]{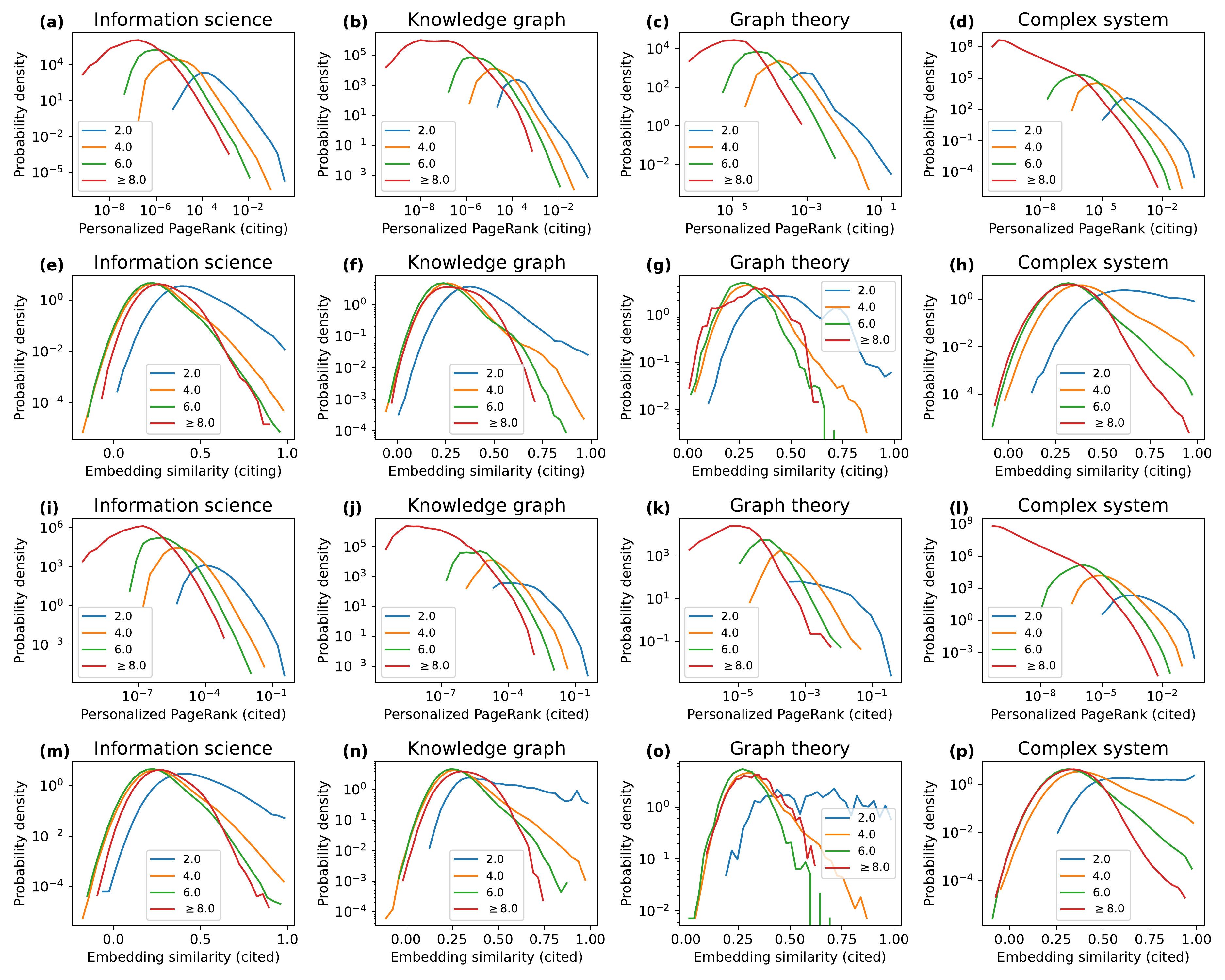}
\caption{The probability density of generalized relatedness between nodes by the distance in the node split network. The colors correspond to the distances between each pair of nodes. A distance of 2 indicates that they have a link in the original bibliographic coupling or co-citation network, while even distances exceeding 2 indicate that they are not linked in the original BC/CC network.}\label{fig:dist_by_pathlen}
\end{figure*}

Until now, we mainly accounted for the node pairs for which the original relatedness can be estimated; these nodes have second-neighbor relationships with distance $2$ in the node split network. One notable advantage of the proposed method is the computability of the similarity between nodes that are far apart. A logical step forward is to determine the role of such distant node pairs whose distance is $>2$ in the node split network. In other words, it is necessary to investigate node pairs that do not share neighbors. 

A question arises from the above results: do PPR and EMB capture essential node pairs that are neglected in the original relatedness network? We tested this question by comparing the distribution of similarity based on the distance between pairs of nodes. Despite slight differences in disciplines and measures, we commonly found high similarity scores for the nearest node pairs (Figure~\ref{fig:dist_by_pathlen}). Specifically, the PPR $\gtrsim10^{-5}$ for the nearest node pairs; this was much higher than the lower bound of the distant node pairs. The overall range of similarities decreased as the distance increased, indicating that BC and CC mainly captured significant node pairs. As an illustrative example, the average PPR similarity was $\simeq 0.00078$ (nearest), $\simeq 3.86 \times 10^{-5}$ (distance of 4), $\simeq 7.528 \times 10^{-6}$ (distance of 6), and $\simeq 1.40 \times 10^{-6}$ (distance greater than 6) for information science. However, we also address the fact that the distributions overlap widely with each other. For information science, the maximum PPR similarity among the distant node pairs was $\simeq 0.08147$, which can be considered as significant; there were only $320$ pairs among $1\,964\,258$ nearest node pairs exceeding the maximum PPR similarity of the distant node pairs ($\simeq 0.08147$). This pattern was also observed for the other disciplines (Table~\ref{tab:signodes}). Therefore, we conclude that BC and CC failed to capture some of the node pairs that should be considered regarding PPR.

We continue our analysis with EMB, which is a low-cost alternative to generalized similarity. Similar to PPR, the EMB for the nearest node pairs displays high similarity scores with the heavier right tail compared with the distant node pairs, and the overall similarity for node pairs seems to become smaller with distance (Figure~\ref{fig:dist_by_pathlen}). However, this trend is not always valid for nodes that are far apart. The average EMB similarity between citing layers for information science is $\simeq 0.4387$ (nearest), $\simeq 0.2488$ (distance of 4), and $\simeq 0.2295$ (distance of 6); however, it increases to $\simeq 0.2826$ for node pairs with a distance greater than 6. This trend is similar to that observed in the other three disciplines. In contrast, BC and CC also failed to capture the nodes with the highest EMB similarity (Table~\ref{tab:signodes}). Only a few nearest node pairs show an EMB similarity greater than that of the distant node pairs. The fraction of such nearest pairs is at most $10.40\%$ for the CC of the graph theory papers (see Table~\ref{tab:signodes}). In summary, if we consider only the nearest node pairs, it is easy to neglect the essential node pairs regarding both PPR and EMB; thus, the distant node pairs should also be considered to draw an accurate map of science and technology.

\begin{table}[]
\centering
\caption{The total number of node pairs of the original relatedness and those with the significant similarity. The significant node pairs are the neighbor node pairs in the relatedness network that exceed the maximum PPR similarity among the distant node pairs.}\label{tab:signodes}
\begin{tabular}{llllllll}
\hline
\multirow{2}{*}{\textbf{Similarity}} & \multirow{2}{*}{\textbf{Field of Study}} & \multicolumn{2}{l}{\textbf{\begin{tabular}[c]{@{}l@{}}Total\\ node pairs\end{tabular}}} & \multicolumn{2}{l}{\textbf{\begin{tabular}[c]{@{}l@{}}Significant \\ node pairs\end{tabular}}} & \multicolumn{2}{l}{\textbf{Fraction (\%)}} \\
 &  & \textit{BC} & \textit{CC} & \textit{BC} & \textit{CC} & \textit{BC} & \textit{CC} \\ \hline
\multirow{4}{*}{\textbf{PPR}} & \textit{Information science} & 1\,964\,258 & 888\,130 & 320 & 840 & 0.016 & 0.095 \\
 & \textit{Knowledge graph} & 900\,150 & 49\,714 & 485 & 713 & 0.054 & 1.43 \\
 & \textit{Graph theory} & 47\,908 & 1500 & 271 & 248 & 0.57 & 16.53 \\
 & \textit{Complex system} & 1\,120\,910 & 136\,688 & 3\,987 & 4\,048 & 0.36 & 2.96 \\ \hline
\multirow{4}{*}{\textbf{EMB}} & \textit{Information science} & 19\,64\,258 & 888\,130 & 76 & 150 & 0.0038 & 0.016 \\
 & \textit{Knowledge graph} & 900\,150 & 49\,714 & 474 & 236 & 0.052 & 0.47 \\
 & \textit{Graph theory} & 47\,908 & 1\,500 & 388 & 156 & 0.81 & 10.40 \\
 & \textit{Complex system} & 1\,120\,910 & 136\,688 & 26 & 20 & 0.0023 & 0.014 \\ \hline
\end{tabular}
\end{table}

\subsection{Impact of sampling methods for the generalized relatedness}\label{sec:sampling}

\begin{figure*}[!ht]
\centering
\includegraphics[width=\textwidth]{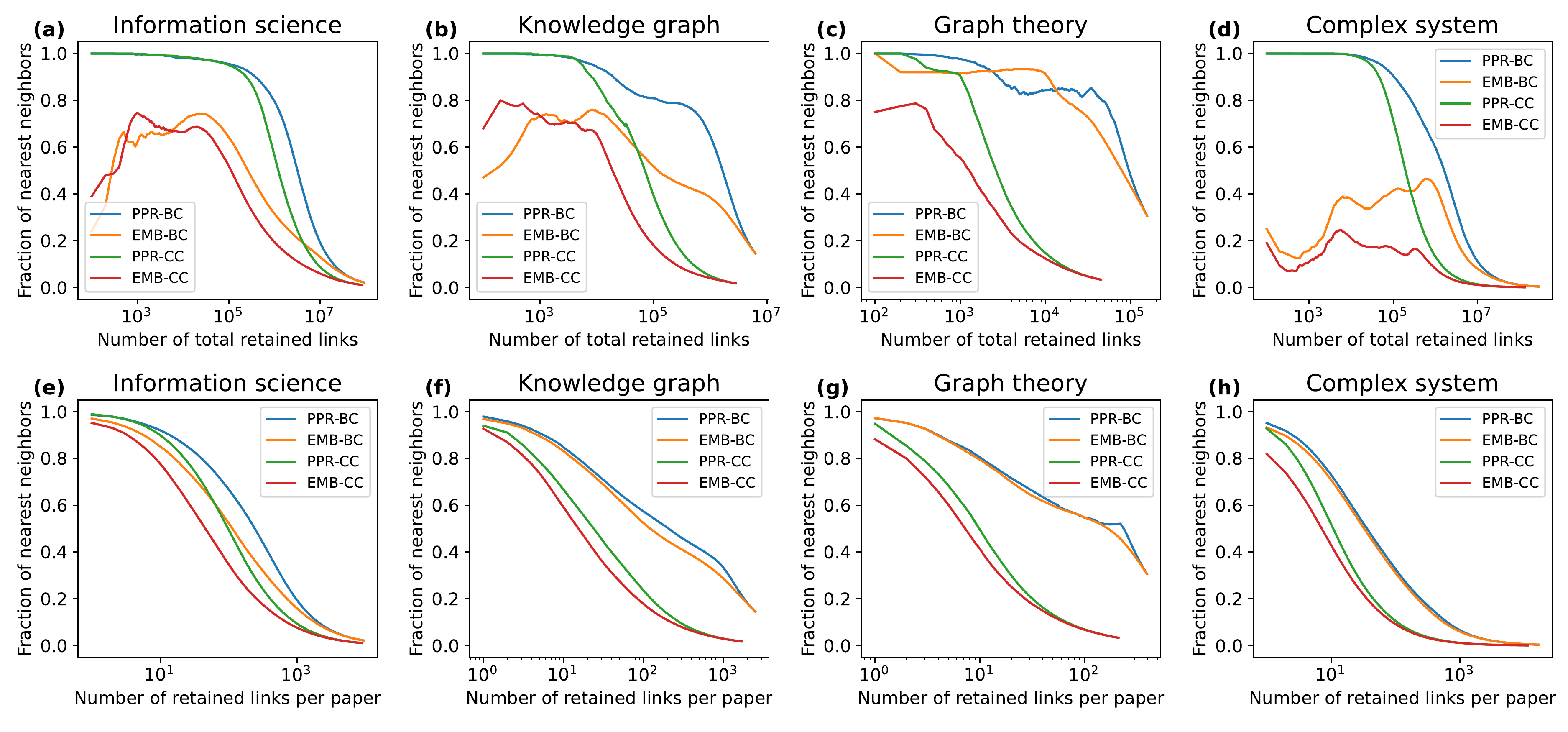}
\caption{Fraction of corresponding BC/CC links that are within a certain number of samples with the highest generalized relatedness similarities: (a--d) We sample the links globally so that only the top $k$ strongest relations in the network are kept. (e--h) We sample the links locally so that only the top $k$ strongest relations per paper remain. For both samplings, we randomly select a link for ties.}\label{fig:sampling}
\end{figure*}

In the previous section, we showed that BC and CC may not capture all significant node pairs regarding generalized relatedness. A natural step forward is to find the proper way to search for important interrelations. For a large-scale analysis that may consist of millions of publications, using the entire set of relations incurs a high computational cost in terms of both time and memory. Sometimes, this task is beyond the limits of modern computers. As an illustrative example, there are approximately $70\,000\,000$ journal papers in the Microsoft Academic Graph. Our generalization computes the relatedness for all node pairs with directions, and thus, there are $\sim 4.9\times10^{15}$ link weights in total. If a single weight is stored in a $64$-bit float, the total memory usage is $\simeq 3.92\times10{16} = 3.92$ petabytes, excluding the overhead of data structures. This is certainly beyond the capabilities of modern computers. In some cases, information on the strongest relations is more important than other information. If one uses relatedness to recommend related nodes from a single node, it is not necessary to obtain the similarity with all other nodes. Instead, it is necessary to calculate the similarity of the top related nodes \citep{lofgren2016personalized}. 

One conventional approach is to handle the problem with link filtering based on the significance of links, that is, to only retain updates on significant links. The ultimate goal of this filtering is clear; the filtering needs to extract the links with relevant information while preserving the key features that should be highlighted. However, the highlighted features are all different from each other. There are two perspectives on link filtering: global and local filtering. Global filtering removes links based on the global rank, which is measured based on all links and nodes \citep{yu2018identifying}. This approach aims to retain significant links to maintain the global structure of the network, for example, backbone extraction \citep{serrano2009extracting}. On the other hand, local filtering uses the rank measured for a single node. As an illustrative example, we can extract the most significant link for each node \citep{boyack2011clustering}. Sometimes, the number of filtered links is based on information entropy \citep{lee2010googling}.

We tested both perspectives using simple filtering based on local and global rankings. We kept the links with weights ranked in the top $k$ compared with all other links in the dataset to test the global filtering. The rest were dismissed. For the local filtering approach, only the top $r$ strongest links per node were retained. We then counted the number of nearest node pairs in the original relatedness network by sample size. We hypothesize that the fraction of the nearest neighbor should be decreased by increasing the number of remaining links in the generalized relatedness network. This hypothesis is based on our findings in Sections~\ref{subsec:similarity}~and~\ref{subsec:missing_links}. We found missing essential links in the original BC/CC network regarding generalized relatedness. However, we also found that there were strong correlations between the generalized relatedness and the original relatedness (BC and CC, see Figure~\ref{fig:scatterplot}); thus, the strongest links of the original relatedness were also the strongest links in the generalized relatedness network. In addition, the range of overall similarities decreased as the distance increased (see Figure~\ref{fig:dist_by_pathlen}). Thus, as many links remain, the fraction of the strongest links should gradually decrease.

Indeed, we found that the fraction in the PPR network was a decreasing function of the number of retained links for both global and local filtering, except for the case of graph theory (see Figure~\ref{fig:sampling}). In the graph theory case, there was a range in which the fraction increased to between $\sim 10^4$ and $\sim10^5$ nodes. Otherwise, PPR was a stable decreasing function of the retained links. In contrast, EMB exhibited unexpected instability for global filtering. The fraction in the EMB network was increased by the number of retained links when it was small (see Figure~\ref{fig:sampling}). Thus, global filtering did not capture significant node pairs. Instead, local filtering showed a monotonically decreasing function with the number of retained links, as we hypothesized. In the entire range of the number of retained links, PPR showed a higher fraction of nearest neighbors compared with EMB, which is consistent with our findings regarding the correlation between the generalized and original relatedness (see Table~\ref{tab:corr}). In summary, local sampling is stable for both PPR and EMB, yet global sampling is only stable for PPR. 

\section{Discussion}
\label{sec:discussion}
In this study, we expanded the concept of a node split network to yield similarities from the generalized perspective of BC and CC. We then tested our approach using an empirical dataset of scientific papers from the fields of information science, knowledge graphs, graph theory, and complex systems. With these empirical datasets, we demonstrated the merits of our new approach. First, PPR can estimate the similarity between all node pairs that can reach each other. Moreover, EMB does not require reachability and can be calculated for any node pairs in the dataset. This approach was inspired by the coupling relatedness of BC and CC; thus, it can be considered a generalization of the relatedness measures. The original measures can only be calculated for nodes that have common neighbors; however, our proposed measures do not have such limitations. The proposed method showed a high similarity to the original coupling relatedness. PPR showed better similarity with the original relatedness and stability with the sampling, but it is expensive to calculate. Alternatively, EMB can be used as a low-cost option for generalized relatedness without suffering from a massive computational load with some trade-offs for accuracy. EMB is sensitive to sampling methods, and therefore, it should be used with local sampling. Our validation sets were rather small subsets with thousands of scientific items (papers), which certainly do not represent the entirety of science and technology. The importance of non-detected links in the original relatedness may be smaller for other fields. However, we believe that the results should be consistent with the fields because the proposed method was derived from the innate structure of citation networks. Note that we did not consider the time scale, \textit{e.g.}, the citation time window. A detailed analysis of the time scale will boost our awareness of the structure of science and technology. For example, relatedness may be missing when two articles are published far apart in terms of time; however, we leave this as a future subject to study. 
 
In the modern big-data era, computational complexity is a significant hurdle in scaling up the proposed methods. For example, there are $\simeq250$ million items in the Microsoft Academic Graph. PPR is computationally expensive, even more than the original relatedness. However, this complexity can be reduced by using fast contemporary algorithms depending on the task ~\citep{lofgren2016personalized}. EMB is rather inaccurate; however, it is more computationally efficient than PPR and is still appropriate for use as an alternative similarity. Many studies have examined BC and CC in detail, but most of these studies focused on measuring more accurate similarities between nodes that share common neighbors using citations and other metadata \citep{kessler1963bibliographic, kessler1963an, small1973co, boyack2010co}. Our analysis suggests that these kinds of coupling relatedness may be generalized solely using citations, using a detailed understanding of the citation structure itself. Our study is not limited to a generalization of the BC and CC networks, and arguably provides insight into future scientific studies; the data analysis should be based on a solid comprehension of the structure of citations. If comprehensive understanding is accomplished, we will also achieve unbiased and quantitative insights on the progress of science and technology. 

\section*{Author's contribution}
Jinhyuk Yun: the author confirms sole responsibility for the study design, data collection, analysis and interpretation of results, and manuscript preparation.

\section*{Acknowledgments}
We thank Dr. June Young Lee and Dr. Jinseo Park for their invaluable discussions on this research. The National Research Foundation (NRF) of Korea Grant funded by the Korean Government supported this work through Grant No. NRF-2020R1A2C1100489. The funders had no role in the study design, data collection and analysis, decision to publish, or preparation of the manuscript.

\bibliographystyle{elsarticle-harv}
\bibliography{references}

\end{document}